\documentclass{elsart}

\usepackage{graphicx}

\usepackage{epsfig}

\usepackage{amssymb}

\begin{document}

\begin{frontmatter}

\title{Probability and complex quantum trajectories}

\author{Moncy V. John}

\address{Department of Physics, St. Thomas College, Kozhencherry, Kerala 689641,\\
India.
}

\begin{abstract}
 It is shown that  in the complex trajectory representation of quantum mechanics,  the Born's $\Psi^{\star}\Psi$ probability density  can be obtained from the imaginary part of the  velocity field of particles on the real axis. Extending this probability axiom to the complex plane, we  first attempt   to find a probability density by solving an appropriate conservation equation. The characteristic curves of this conservation equation are found to be the same as the complex paths of particles in the new representation.  The boundary condition in this case is that the extended probability density  should  agree with the quantum probability rule along the real line. For the simple, time-independent, one-dimensional problems  worked out here,  we find that   a conserved probability density can be derived from the velocity field of particles, except in regions where the trajectories were  previously suspected to be nonviable.  An alternative method to find this  probability density in terms of a trajectory integral, which is  easier to implement on a computer and  useful for single particle solutions, is also presented. Most importantly, we show,  by using the complex extension of Schrodinger equation, that the desired conservation equation can be derived from this definition of probability density.

\end{abstract}


\end{frontmatter}

\section{Introduction}

In a previous work  \cite{mvj}, the quantum Hamilton-Jacobi equation [2-5] (QHJE) was  made use of   to demonstrate the existence of particle trajectories  in a complex space, for different quantum states. This complex quantum trajectory  representation was obtained by modifying the de Broglie-Bohm (dBB) approach to quantum mechanics \cite{dBB}, which allows particle motion guided by the wave function. One of the  advantages of the resulting theory, which offers a new interpretation of quantum mechanics, is that it does not face the  problem of stationarity of particles in bound states, encountered in the dBB representation. Another trajectory approach to quantum mechanics,  which also claims the  absence of this problem,  is the  representation developed by Floyd, Faraggi, Matone (FFM) and  others [7-9]. 

The new complex trajectory representation proceeds by first substituting $\Psi=e^{i\hat{S}/\hbar}$ in the Schrodinger equation to obtain the quantum Hamilton-Jacobi equation 

\begin{equation}
\frac {\partial \hat{S}}{\partial t} + \left[ \frac{1}{2m}\left( \frac
{\partial  \hat{S}}{\partial x}\right)^2 +V\right]  =
\frac{i\hbar}{2m} \frac{\partial^2 \hat{S}}{\partial x^2}, \label{eq:qhje}
 \end{equation}
and then postulating an equation of motion

\begin{equation}
m\dot{x} \equiv \frac {\partial \hat{S}}{\partial x}= \frac {\hbar
}{i} \frac {1}{\Psi}\frac {\partial \Psi}{\partial x} 
\label{eq:xdot}   \end{equation}
for the particle.  The trajectories $x(t)$ of the particle in the complex $x$-plane  are obtained by integrating this equation with respect to time \cite{mvj}. It was observed that the above identification $\Psi=e^{i\hat{S}/\hbar}$  helps to utilize all the information contained in $\Psi$ while obtaining the trajectory. (The dBB approach, which uses  $\Psi=R e^{i{S}/\hbar}$ does not have this advantage.) The complex eigentrajectories in the free particle, harmonic oscillator and potential step problems and complex trajectories for a wave packet solution were obtained in \cite{mvj}. The representation was  extended to three dimensional problems, such as the hydrogen atom,  by Yang \cite{yang} and  was used to investigate one dimensional scattering problems and bound state problems by  Chou and Wyatt \cite{wyatt,wyatt2}. Later, a complex trajectory
approach for solving the QHJE was developed  by Tannor and
co-workers \cite{goldfarb}. The  QHJE was derived independently by Sanz and Miret-Artes \cite{sanz}, who also found the complex trajectory representation  useful in  better understanding the  nonlocality in quantum mechanics \cite{sanz1,sanz2}. 

 It is  well known that the QHJE as given in Eq. (\ref{eq:qhje})  was used by many physicists such as Wentzel, Pauli and Dirac,  even during the time of inception of quantum mechanics \cite{oldpaps}. In a commendable work in 1982, Leacock and Padget \cite{leacock} have used the QHJE to obtain eigenvalues in many bound state problems, without actually having to solve the corresponding Schrodinger equation. However,  there were no trajectories in  their work and it was only  in \cite{mvj} that the equation of motion (\ref{eq:xdot})  explicitly solved and the complex trajectories of particles in any quantum state obtained and drawn,  for the first time. This paper also  highlighted the interpretational value of the complex quantum trajectory representation, vis-a-vis the Bohmian mechanics. Eventhough this  formulation requires that the wavefunction is known from a separate calculation, it is generally conceded that it was the work in \cite{mvj} which provided a complex  trajectory interpretation of quantum mechanics \cite{sanz2,goldfarb2}.  
\medskip

Instead of computing the complex trajectories $x(t)$, the complex paths  $x_i(x_r)$ in the above scheme can directly be found  by integrating the equation

\begin{equation}
\frac{dx_i}{dx_r}=\frac{\dot{x}_i}{\dot{x}_r}, \label{eq:path}
\end{equation}
where Eq. (\ref{eq:xdot}) shall be used. In \cite{mvj}, it was noted that even  for an eigenstate, the particle can be in any one of its infinitely many possible quantum trajectories, depending on its initial position in the complex plane. Therefore,   the expectation values of dynamical variables are to be evaluated over an ensemble of particles in all
possible trajectories. It was postulated that the average of a dynamical variable
$O$ can be obtained using the  measure $\Psi^{\star} \Psi $  as

\begin{equation}
<O> = \int_{-\infty}^{\infty} O \Psi^{\star} \Psi
 \;  dx, \label{eq:mean}
\end{equation}
where the integral is to be taken along the real axis  \cite{mvj}. Also it was noted that in this form, there is no need to make the conventional operator replacements. The above postulate is equivalent to the Born's probability axiom for observables such as position, momentum, energy, etc., and one can show that 
$<O>$ coincides with the corresponding quantum mechanical expectation
values. This makes the new scheme  equivalent to standard 
quantum mechanics when  averages of dynamical variables are computed.

One of the  challenges before this complex quantum trajectory representation, which is an ontological theory of particle motion,  is to explain the  quantum probability axiom. In the dBB approach, there were several attempts to obtain the $\Psi^{\star}\Psi$ probability distribution from more fundamental assumptions \cite{dBB}. In the present paper, we first attempt to obtain this distribution along the real line from  the velocity of particles in the complex trajectory representation. It is found that always there exists a direct relationship between the $\Psi^{\star}\Psi$  distribution and the imaginary component of the particle's velocity on the real line. Since this distribution is defined and used only along the real axis, the conservation equation for probability in the standard quantum mechanics is valid here also, without any modifications.

At the same time, since we have the complex paths,  it would be natural to consider the probability for the particle to be  in a particular path. In addition, we may consider the probability to find the particle around different points in the same path, which can also be different. Thus it is desirable to extend the probability axiom  to the  $x_rx_i$-plane. But in this case,  it becomes  necessary to see whether probability conservation holds everywhere in the plane. A recent paper by Poirier \cite{poirier} addresses this issue and obtains some negative results for the choices made for such a distribution. Poirier tries to define an extended probability by working with $\Psi$, the solution of time-dependent Schrodinger equation. The probability density at a point $x$ in the complex plane is postulated as $\rho (x)= \bar{\Psi (x)}\Psi (x)$ where $\bar{\Psi (x)} \equiv \Psi^{\star}(x^{\star})$. The complexified flux is chosen as $j(x,t)=v(x,t) \rho (x,t) =-(i\hbar /m) \Psi^{\star}(x^{\star})\Psi^{\prime}$ where $v(x,t)\equiv \dot{x}$ is given by equation (\ref{eq:xdot}) and prime denotes spatial differentiation. With the help of time-dependent Schrodinger equation, the author shows that, in general, 

$$
\frac{\partial \rho}{\partial t}\neq j^{\prime}(x,t).
$$
This, arguably leads to nonconservation of probability along trajectories. But it shall be reminded that this negative result is based  on the choices made in \cite{poirier} for the probability density and flux.

In the present paper, on the other hand, we first show that the Born's probability density, defined along $x=x_r$, is obtainable as the exponential of an integral over the real line, of $\dot{x}_i$, the imaginary part of the particle's complex velocity. Next, extending this probability axiom to the entire $x_rx_i$-plane, we attempt to  solve the appropriate  equation for  a  conserved probability density. It is also demanded that this quantity should  agree with Born's probability rule along the real axis, which is the boundary condition in this case. An important result  obtained is that the characteristic curves of the conservation equation, along which the information about the solution propagates, are the same as the  paths of particle  in the  complex trajectory representation. We find that  there exist  conserved probability densities which agree with the boundary condition in most of the examples  considered. An alternative method to evaluate this in terms of a trajectory integral, in a manner similar to obtaining the $\Psi^{\star}\Psi$ probability distribution along the real line, is also presented.  This latter method shall be of interest while solving the QHJE for the motion of individual particles. Most importantly, we show that  the above conservation equation can be derived from this definition of probability, by using the complex extension of Schrodinger equation.

\section{Probability from velocity field}

 Let us recall that  in this scheme, the real part of the velocity of a particle  on  the real line, denoted as $\dot{x}_r (x_r,0)$, always agrees with the velocity of particle in the dBB representation  \cite{mvj}. To obtain the quantum probability function from the velocities, which is our first goal, we   note further that the imaginary component $\dot{x}_i$ of the velocity of the particle in the complex trajectory can be written as

\begin{equation}
\dot{x}_i = -\frac {\hbar }{2m} \frac {\left[ \Psi ^{\star }\frac
{\partial  \Psi  }{\partial x_r}+(\frac {\partial \Psi} {\partial
x_r})^{\star}\Psi \right]} {\Psi ^{\star} \Psi}. \label{eq:xidot}
\end{equation}
 This helps to write the   probability density to find the particle around some point $x=x_r$ as

\begin{equation}
\Psi^{\star}\Psi (x_r,0) \equiv P(x_r) ={\cal N} \exp \left({-\frac{2m}{\hbar}\int^{\dot{x_r}} \dot{x}_i dx_r}\right) , \label{eq:psistarpsi}
\end{equation}
where the integral is taken along the real axis.   This possibility of regaining the quantum probability distribution from the velocity field is a unique feature of the complex trajectory formulation. For instance, in the dBB approach, the velocity fields  for all bound eigenstates are zero everywhere  and it is not possible to obtain  a relation between velocity and probability. The FFM trajectory representation, on the other hand, does not claim any connection with probability.

In the following, we limit ourselves to one dimension and to time-independent problems. The variable $x$ is always assumed to be complex.
As stated earlier, in this new representation, which is based on the existence of trajectories in the complex plane,  even a particle in an eigenstate can be in any one of the infinitely many possible trajectories, depending on its initial position in the complex plane. Thus it is also desirable to  look for the probability of a particle to be  in an area $dx_rdx_i$ around some point ($x_r, x_i$) in the complex plane. Let this quantity be denoted as $\rho(x_r, x_i)dx_rdx_i$.

An explicit expression for   $\rho(x_r,x_i)$ can be arrived at in the following way. Extending the probability density to the entire $x_rx_i$-plane demands a conservation equation of the form (for time-independent cases)

\begin{equation}
\frac{\partial (\rho \dot{x}_r)}{\partial x_r}+\frac{\partial (\rho \dot{x}_i)}{\partial x_i}=0. \label{eq:cons}
\end{equation}
To solve this partial differential equation, we may write  $\rho(x_r,x_i)=h(x_r,x_i) f(p)$, where $h(x_r,x_i)$ is some solution of Eq. (\ref{eq:cons}) and $p$ is some combination of $x_r$ and $x_i$, whose value remains a constant along its characteristic curves  \cite{riley}. Substituting this form of $\rho$ into  (\ref{eq:cons}), we see that  the characteristic curves are obtained by integrating the equation

\begin{equation}
\frac{dx_r}{\dot{x}_r}=\frac{dx_i}{\dot{x}_i} \hspace{1cm} \hbox{or} \hspace{1cm} \frac{dx_i}{dx_r}=\frac{\dot{x}_i}{\dot{x}_r}, \label{eq:diff}
\end{equation}
 which is found to be the same as Eq. (\ref{eq:path}). This demonstrates the important property that the characteristic curves  for the above conservation equation are  identical to the complex paths of particles in the present quantum trajectory representation. 

We may now  find the exact form of $f(p)$ by requiring that $\rho(x_r, 0)$ agrees with the probability $P(x_r)$, which is  the boundary condition in this case. Let the integration constant in the above equation (\ref{eq:diff})  be the (real) coordinate of any one point of crossing of the trajectory on the real axis,  denoted as $x_{r0}$. Since the characteristic curves are identical to the complex paths,  one can take $x_{r0}$ as the  constant $p$ along the characteristic curve and let it be expressed in terms of $x_r$ and $x_i$. The assumed  form for  the extended probability distribution $\rho$ may then be written as

\begin{equation}
\rho(x_r,x_i) = h(x_r,x_i) f(x_{r0}). \label{eq:rhogen}
\end{equation}
Now we can choose  $f(x_{r0})$ subject to the  boundary condition. At the point $x=x_{r0}$ at which the curve $C$ crosses the real line, we demand (the boundary condition)

\begin{equation}
\rho(x_{r0},0) = h(x_{r0},0) f(x_{r0}) =P(x_{r0})  \label{eq:fxr0}
\end{equation}
and obtain $f(x_{r0})$. Expressing $x_{r0}$ in terms of $x_r$ and $x_i$ in $f(x_{r0})$,  Eq. (\ref{eq:rhogen}) gives $\rho(x_r,x_i)$.

A word of caution is appropriate here. There may be instances, as we shall see below, when the boundary condition overdetermines the problem and we are unable to find a solution. It is observed that this happens in certain regions of the complex space where the trajectories were previously suspected to be nonviable. 

We may also note that  the differential in Eq. (\ref{eq:path}), given by

\begin{equation}
\dot{x}_idx_r - \dot{x}_rdx_i =0
\end{equation}
is inexact because 

\begin{equation}
\frac{\partial \dot{x}_r}{\partial x_r}\neq -\frac{\partial \dot{x}_i}{\partial x_i},
\end{equation}
unless both the partial derivatives are zero. This is due to the Cauchy-Riemann conditions satisfied by the analytic function $\dot{x}$ in the complex $x$-plane (except at its singular points). However, this inexact differential can always be made exact by multiplying with an integrating factor $\mu (x_r, x_i)$, which obeys

\begin{equation}
\frac{\partial (\mu \dot{x}_r)}{\partial x_r}= -\frac{\partial (\mu \dot{x}_i)}{\partial x_i}.
\end{equation}
Thus we see that the integrating factor $\mu$ can serve as $h(x_r,x_i)$ in Eq. (\ref{eq:rhogen}).

The following observation  may be helpful in finding $h(x_r,x_i)$, or even $\rho(x_r,x_i)$,  in the case of some special potentials.  Let us denote $d\Psi /dx \equiv \chi(x)$. Then using the time-independent Schrodinger equation, one can rewrite the equation of motion (\ref{eq:xdot}) as 

\begin{equation}
\dot{x}=\frac{2i(E-V)}{\hbar}\frac{\chi}{\chi \;^{\prime}} \label{xdot2},
\end{equation}
where $\chi\;^{\prime}\equiv d\chi /dx$. For $V=V_0$, a real constant, one can integrate this to obtain

\begin{equation}
\chi(x)=A\;\exp\left[{\frac{2i(E-V_0)}{\hbar}t}\right],
\end{equation}
 from which we get the trajectory of the particle in the complex $x$-plane as $\chi^{\star}(x)\chi (x)=\mid A\mid^2$. Moreover, one can see that

\begin{equation}
(\chi\;^{\prime})^{\star}\chi\;^{\prime}=\frac{4(E-V_0)^2}{\hbar^2}\frac{\chi^{\star}\chi}{\dot{x}^{\star}\dot{x}}. \label{eq:chisq}
\end{equation}
But for constant potentials, $(\chi\;^{\prime})^{\star}\chi\;^{\prime} \propto \Psi^{\star}\Psi$ in the complex plane. It can be seen that the above expression (\ref{eq:chisq}) satisfies the conservation equation  and agrees with the boundary condition. In other words, the extended, conserved probability $\rho(x_r,x_i)$ in this case can be written as

$$
\rho(x_r,x_i) = \Psi^{\star}\Psi =\frac{\hbar^2}{m^2} \frac{\mid A\mid^2}{ \dot{x}^{\star}\dot{x}},
$$
and hence $\rho$ varies inversely as $\mid \dot{x}\mid^2$, as the particle moves along a particular trajectory with fixed $A$. This result is not contradictory to the WKB result that $\Psi^{\star}\Psi \propto 1/v_{\hbox{classical}}$ for constant potentials \cite{sakurai}, since $\mid \dot{x}\mid $ along a trajectory in our case is not the same as  $v_{\hbox{classical}}$.

\medskip

In the case of harmonic oscillator potential, we shall see below that an expression of the form ${\mid A\mid^2}/({ \dot{x}^{\star}\dot{x}})$ will give the solution $h(x_r,x_i)$ of the conservation equation, but for the particular solution which agrees with the boundary  condition, one need to find $f(x_{r0})$ too.

\section{Examples}

Let us now first apply the procedure to the harmonic oscillator. In the $n=0$ case, we have $\dot{x}_i=(\hbar \alpha^2/m)x_r$  and the quantum probability density is regained by using Eq. (\ref{eq:psistarpsi}) as

\begin{equation}
P(x_r) \equiv {\cal N} \exp\left (-\frac{2m}{\hbar}\int^{x_{r}} \dot{x}_i dx_r\right)={\cal N}_0e^{-\alpha ^2x^{2}_r}.
\end{equation}
To find $\rho(x_r,x_i)$, first let us note that $h=h_0$, a constant, can be a solution of the conservation equation (\ref{eq:cons}) in this case. Then we find $f(x_{r0})$  using Eq.(\ref{eq:fxr0}) as proportional to $\exp(-\alpha ^2x^{2}_{r0})$. To put $x_{r0}$ in terms of $x_r$ and $x_i$,  Eq. (\ref{eq:path}) shall be integrated to obtain the paths as $x_r^2+x_i^2=\hbox{constant}$ \cite{mvj}. Equating this constant to $x_{r0}^2$, we can finally write

\begin{equation}
\rho(x_r,x_i) \propto e^{-\alpha^2(x_r^2+x_i^2)}.
\end{equation}
This conserved probability is plotted in Fig. 1.

Similarly in the $n=1$ eigenstate of this case, one  uses Eq. (\ref{eq:psistarpsi}) to regain $P(x_r) ={\cal N}_1 x_r^2\exp(-\alpha^2x_r^2)$. In the next step, $h(x_r,x_i)=(x_r^2+x_i^2)$ is found to be a solution of the conservation equation (\ref{eq:cons}). (One could guess this expression from the form of $\dot{x}^{\star}\dot{x}$ in this case, as it appears in its denominator.)  $f(x_{r0})$ can now be found as

\begin{equation}
f(x_{r0})  \propto e^{-\alpha^2x_{r0}^2}.
\end{equation}
The complex paths in this case are given by $(\alpha^2x_r^2-\alpha^2x_i^2-1)^2+4\alpha^4x_r^2x_i^2 = \; \mid A\mid^2$, a constant. (Note that also this  appears in the expression for $\dot{x}^{\star}\dot{x}$. Explicitly, one obtains $\dot{x}^{\star}\dot{x} =\mid A\mid^2/(x_r^2+x_i^2)$). Equating this constant to $(\alpha^2x_{r0}^2-1)^2$, one can obtain $f(x_{r0})$ and also the extended  probability density $\rho$, in terms of $x_r$ and $x_i$. But while taking square roots, one need to be careful. It shall be noted that for the region containing the subnests in the harmonic oscillator  (with $A<1$ in the present $n=1$ case) \cite{mvj}, the boundary condition overdetermines the problem, resulting in there being no solution. But for the region outside it, one can write

\begin{equation}
f(x_{r0}) \propto \exp ({- \sqrt{(\alpha^2x_r^2-\alpha^2x_i^2-1)^2+4\alpha^4x_r^2x_i^2 }}\;),
\end{equation} 
and therefore,

\begin{equation}
\rho(x_{r},x_i)\propto (x_r^2+x_i^2)\exp ({-\sqrt{(\alpha^2x_r^2-\alpha^2x_i^2-1)^2+4\alpha^4x_r^2x_i^2 }}\;) \label{eq:n1prob}
\end{equation}
which is plotted in Fig. 2.

Now let us consider another example of a particle in an infinite square well potential, between $x=0$  and $x=a$. Using the energy  eigenfuntion $\psi_n=\sqrt{{2}/{a}} \sin (n\pi x/a)$, one can find that the complex paths are given by

\begin{equation}
 \cosh \left(\frac{2n\pi x_i}{a}\right) +\cos \left(\frac{2n\pi x_r}{a}\right) =\mid A\mid^2.
\end{equation}
We can choose $h(x_r,x_i) = \cosh (2n\pi x_i/a)-\cos (2n\pi x_r/a)$ as a solution of the conservation equation. (Here again, we have $\dot{x}^{\star}\dot{x} =\mid A\mid^2/h(x_r,x_i)$.) Interestingly, here, we have $f(x_{r0})=\hbox{ constant}$ and hence, 

\begin{equation}
 \rho(x_r,x_i) \propto \cosh \left(\frac{2n\pi x_i}{a}\right)-\cos \left(\frac{2n\pi x_r}{a}\right) .
\end{equation}
The $n=1$ case of this  distribution is plotted in Fig. 3.

A similar situation arises in the case of complex trajectories in the potential step problem discussed in \cite{mvj}. Again, we find $\dot{x}^{\star}\dot{x} =\mid A\mid^2/h(x_r,x_i)$ and $f(x_{r0})=\hbox{constant}$. For the reflection constant equal to $1/2$, we thus get, for the potential step,

\begin{equation}
 \rho(x_r,x_i) \propto e^{-2kx_i}+\sqrt{2}\cos (2kx_r)+\frac{1}{2}e^{2kx_i},
\end{equation}
which is plotted in Fig. 4.

\section{Trajectory integral method to find $\rho$}

It may now appear that the above two approaches to probability, namely, the one which gives the Born's $\Psi^{\star}\Psi$ probability along the real line using equation (\ref{eq:psistarpsi}) and the other, which gives a conserved probability density over the entire $x_rx_i$-plane by solving a conservation equation, do not have anything in common. But in this section, we show that there is a more elegant way of obtaining 
$\rho(x_r,x_i)$, by uniting them. This method may also be useful in finding the variation of probability along any trajectory.

First, we postulate that if $\rho_0$, the extended probability density at some point $(x_{r0},x_{i0})$ is given, then $\rho(x_r,x_i)$ at another point that lies on the trajectory which passes through $(x_{r0},x_{i0})$, is

\begin{equation}
\rho (x_r,x_i) = \rho_0  \exp\left[  \frac{-4}{\hbar}\int_{t_0}^{t} Im\left(\frac{1}{2}m\dot{x}^2+V(x)\right)dt^{\prime}    \right]. \label{eq:rho_def}
\end{equation}
Here, the integral is taken along the trajectory $[x_r(t^{\prime}),x_i(t^{\prime})]$. One can derive the continuity equation by using the extended version of the Schrodinger equation, which gives

 \begin{equation}
Im(E) = Im\left( \frac{1}{2}m\dot{x}^2+V(x)+\frac{\hbar}{2i}\frac{\partial \dot{x}}{\partial x}\right) =0,
\end{equation}

since energy and time are assumed real \cite{mvj}. This helps to write the above definition (\ref{eq:rho_def}) as

\begin{equation}
\rho(x_r,x_i)=\rho_0 \exp \left( -2\int_{t_{0}}^{t}\frac{\partial \dot{x}_r}{\partial x_r}dt^{\prime}\right),
\end{equation}
which in turn gives

\begin{equation}
\frac{d\rho }{dt} =-2\frac{\partial \dot{x}_r}{\partial x_r}\rho = -\left(\frac{\partial \dot{x}_r}{\partial x_r}+\frac{\partial \dot{x}_i}{\partial x_i} \right) \rho.
\end{equation}
The last step follows from the analyticity of $\dot{x}$. This leads to the continuity equation for the particle, as it moves along: i.e.,

\begin{equation}
\frac{\partial \rho}{\partial t} + \frac{\partial (\rho \dot{x}_r)}{\partial x_r}+\frac{\partial (\rho \dot{x}_i)}{\partial x_i}=0. \label{eq:cons_gen}
\end{equation}
 While evaluating $\rho$ with the help of (\ref{eq:rho_def}) above, one needs to know $\rho_0$ at $(x_{r0},x_{i0})$ and if we choose this point as $(x_{r0},0)$, the point of crossing of the trajectory on the real line, then $\rho_0$ may  take the value $P(x_{r0})$ and may be found using  (\ref{eq:psistarpsi}). Here it shall be reminded that the integral in $P(x_{r0})$ is evaluated over the real line only.

To summarize the  alternative method, we note that  the conserved, extended probability density is 

\begin{equation}
\rho (x_r,x_i) \propto \exp\left(-\frac{2m}{\hbar}\int^{x_{r0}}\dot{x}_i dx_r\right) \exp\left[  \frac{-4}{\hbar}\int_{t_0}^{t} Im\left(\frac{1}{2}m\dot{x}^2+V(x)\right)dt^{\prime}    \right], \label{eq:rho_alt}
\end{equation}
with the integral in the first factor  evaluated over the real line and that in the second factor over the trajectory of the particle. This method is easier to implement numerically on a computer.

We can now demonstrate this trajectory integral method by applying it to the $n=1$ harmonic oscillator state. In this case, taking the numerical values of both $\alpha $ and $\omega$ to be unity (which then also applies to $\hbar /m $), it is easily seen that \cite{mvj}

\begin{equation}
x_r =\pm \sqrt{(1+A\cos 2t)\pm\sqrt {1+A^2+2A\cos 2t}}/\sqrt{2} \label{eq:n1xr}
\end{equation}
and 

\begin{equation}
x_i=\frac{A\sin 2t}{2x_r}.\label{eq:n1xi}
\end{equation}

Here one can write $P(x_{r0})=(1+A)\exp(-(1+A))$. Finding $x_r$ and $x_i$ using the above expressions, we have evaluated numerically

\begin{equation}
\rho(x_r,x_i)=P(x_{r0})\exp \left( -4\int_{0}^{t}\frac{x_r x_i}{(x_r^2+x_i^2)^2} dt\right).\label{n1rho_alt}
\end{equation}
for different $A$ and $t$. This was found to be the same extended distribution as that given by Eq. (\ref{eq:n1prob}) and shown in Fig. 2, with $A>1$. More specifically, in this case, since equations (\ref{eq:rhogen}) and (\ref{eq:rho_alt}) must agree, the surface plots of $h(x_r,x_i)=(x_r^2+x_i^2)$ and the expression

$$
(1+A)\exp \left( -4\int_{0}^{t}\frac{x_r x_i}{(x_r^2+x_i^2)^2} dt\right),
$$
with various values of $A$, $t$ but plotted against $x_r$ and $x_i$ evaluated using equations (\ref{eq:n1xr}) and (\ref{eq:n1xi}), must be the same. 
This too is found to be true, which demonstrates that the two methods of evaluating the probability give identical results.

\section{Discussion}

The complex quantum trajectory representation is   worth pursuing mainly for the alternative interpretation it offers to standard  quantum mechanics.   The standard quantum theory is spectacularly successful in explaining all observations made so far and there is no demand for modifications to  the quantum  probability axiom. It is envisaged that the new complex trajectory representation does not differ from standard quantum mechanics, with regard to its predictions of experimental results. This is ensured by accepting the Born's probability axiom as such in the new theory, on the real line. The quantum probability density, though it is not claimed to be perfectly understood, works well in a miraculous manner. In this paper, we   obtain this distribution along the real line from  the velocity field of particles in the complex trajectory representation. Its conservation on the real line is guaranteed in standard quantum mechanics by virtue of the Schrodinger equation and there is no need for any separate proof for the same in the complex trajectory representation. What we try to do next is to extend it to the complex plane, where it should obey conservation laws, as in the case of any other probability function. Our great expectation  is to find such a probability density in the  complex plane, in the hope of it giving us a clue to the basis of this  distribution.

Here we have first adopted the strategy of solving the conservation equation in the  $x_rx_i$-plane, with the velocity field given by Eq. (\ref{eq:xdot}), applicable to one dimensional, time-independent single particle states.  In this endeavor,  there is a significant improvement in our understanding of the origin of quantum probability. First, we have noticed that the characteristic curves of the conservation equation, along which the information about the solution propagates, are the same as the  paths of particle  in the  complex trajectory representation. The  most notable results we have obtained are that  an extended, conserved probability $\rho(x_r,x_i)$, which agrees with the quantum probability rule along the real line, can exist and that it can be derived from the velocity field of particles. There are  parts of the plane in some examples where such a  probability  cannot be found. But in those cases where it is possible,  we find that $\rho(x_r,x_i)$ can  be written as a product of two factors; one [$f(x_{r0})$], a constant for the given complex path and the other [$h(x_r,x_i)$], dependent on the velocity of the particle as it moves along this path. In particular, we note that the Born's probability rule $\Psi^{\star}\Psi \equiv P(x_r)$ itself, which is valid on the real line, can be written as a product as stated above. On the other hand, in an alternative method of evaluating the extended probability,  $\rho(x_r,x_i)$ is found to be obtainable as $P(x_{r0})$ times an exponential factor which involves  an integral with respect to time along the trajectory, of the imaginary part of the sum of the complex kinetic  and potential energies of the particle. The two methods of obtaining the extended probability density were compared and found to give identical results. The latter trajectory integral method is  easier to implement using a computer and may be particularly useful while solving the QHJE for single particles.

The solutions we obtain for the complex plane are not the same as $\Psi^{\star}\Psi$, except in the case of those constant potentials, discussed in Sec. (4). 
It shall be noted that $\rho$ is defined only over that region of the $x_rx_i$-plane, where characteristic curves which  cross the real axis pass. At all other points, $\rho$ is assumed to be zero. The extended probability may be normalized on this basis. Also  the present representation offers  the viewpoint that a particle is observed  only when it surfaces on the real axis. $P(x_r)$ is the probability for such appearances and shall be used for computations involving measurable quantities. In this way, all the standard results in quantum mechanics continue to be unaffected.

Most importantly, we have seen that the definition (\ref{eq:rho_def}) of the extended probability density leads to the continuity equation (\ref{eq:cons_gen}), while using the complex extension of Schrodinger equation. This definition of $\rho$ can thus be of general validity. But here, $\rho$ is a real function of $x_r$ and $x_i$, which does not obey the Cauchy-Riemann equations, and hence  is not analytic in the complex $x$-plane. It may be noted that in our attempt to have a better understanding of the origin of quantum probability rule, where we could make some inroads, this nonanalyticity does not appear as a  problem.  
Ref. \cite{poirier} strongly advocates an analytic expression for $\rho$, in view of its anticipated advantages in the synthetic time-dependent Schrodinger equation applications. But the definition in \cite{poirier} is shown not to lead to a continuity equation and it is presented there as a negative result.  Whether our definition of extended probability in this paper, which does obey a continuity equation, is helpful for synthetic applications is an open problem, and is worth pursuing.

To conclude, we summarize the positive and negative features of this possible definition of quantum probability density in the complex space, in comparison with the distribution prescribed in \cite{poirier}. First, we see that our definition helps to obtain the Born's $\Psi^{\star}\Psi$ probability density along the real line from the velocity field in the complex trajectory formalism, in a nontrivial way. On the other hand, the distribution in \cite{poirier} is simply the analytic continuation of   $\Psi^{\star}\Psi$ and hence the agreement is trivial. Second, our definition of the probability density for the particle to be around some point in its trajectory, as it moves along, obeys a conservation equation. In contrast, it was shown in \cite{poirier} that the probability density defined in it is not conserved, in general. Furthermore, while solving the conservation equation, we have noted that $\rho$ can be written as a product between two factors, one of which is a constant for a trajectory. This feature was deduced while attempting to  solve the conservation equation and is a  consequence of the fact that trajectories and the characteristic curves of the conservation equation are identical in the complex trajectory representation. This has no parallels in the proposed probability density in \cite{poirier}. A negative aspect of the present formalism is that the probability density is not an analytic function in the complex plane, a feature which is suspected to adversely affect its applicability in the synthetic solutions. This is another feature that distinguishes the present distribution from the analytic one studied in \cite{poirier}. It is noted that in our attempt to understand the origin of quantum probability rule in complex trajectory representation, this nonanalyticity does not pose any  problem. But  the complex probability that appears in \cite{poirier} off of the real axis would be an undesirable feature, and could have posed some problems even if it was found to obey the continuity equation. Thus the  issue of whether analyticity or real-valuedness of probability density is the more desirable feature in synthetic applications can be settled only when one progresses with such applications of the present formalism.

\noindent {\bf Acknowledgments}

It is a pleasure to thank the Referee for valuable suggestions, Professor K. Babu Joseph, Arun and Joe for discussions and the Santhom Computing Facility for hospitality.

\begin{figure}[ht] 
\centering{\resizebox {0.8 \textwidth} {0.8 \textheight }  
{\includegraphics {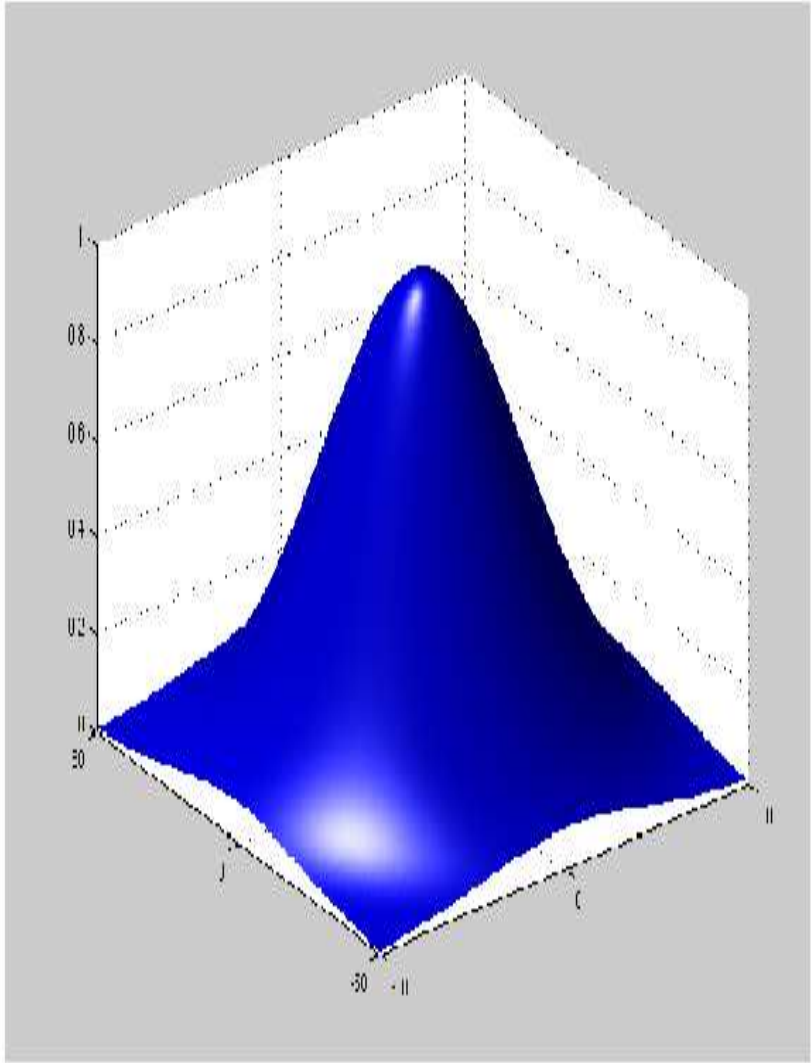}} 
\caption{ The extended probability density $\rho(x_r,x_i)$ for the $n=1$ harmonic oscillator.   }  \label{fig:n0shm}}    \end{figure}

\begin{figure}[ht] 
\centering{\resizebox {0.8 \textwidth} {0.8 \textheight }  
{\includegraphics {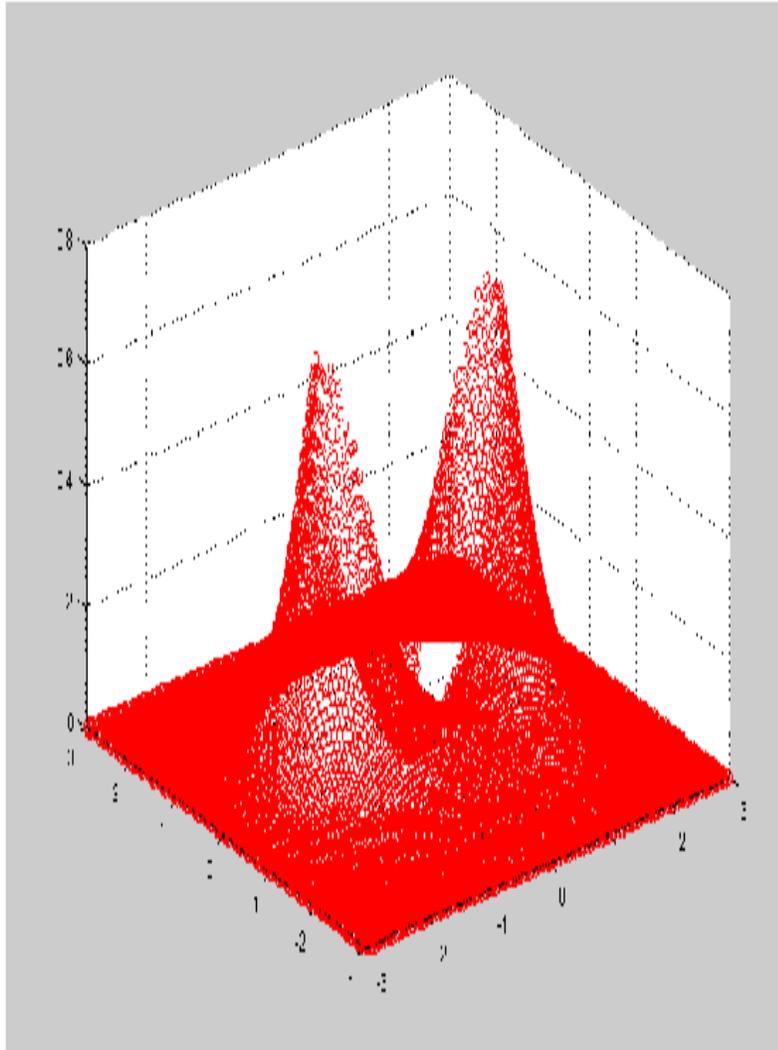}} 
\caption{  The extended probability density $\rho(x_r,x_i)$ for the $n=2$ harmonic oscillator. Here, however, there does not exist a conserved probability for the subnests with $\mid A \mid < 1$ in the interior region.}  \label{fig:n1shm}}    \end{figure}

\begin{figure}[ht] 
\centering{\resizebox {0.8 \textwidth} {0.8 \textheight }  
{\includegraphics {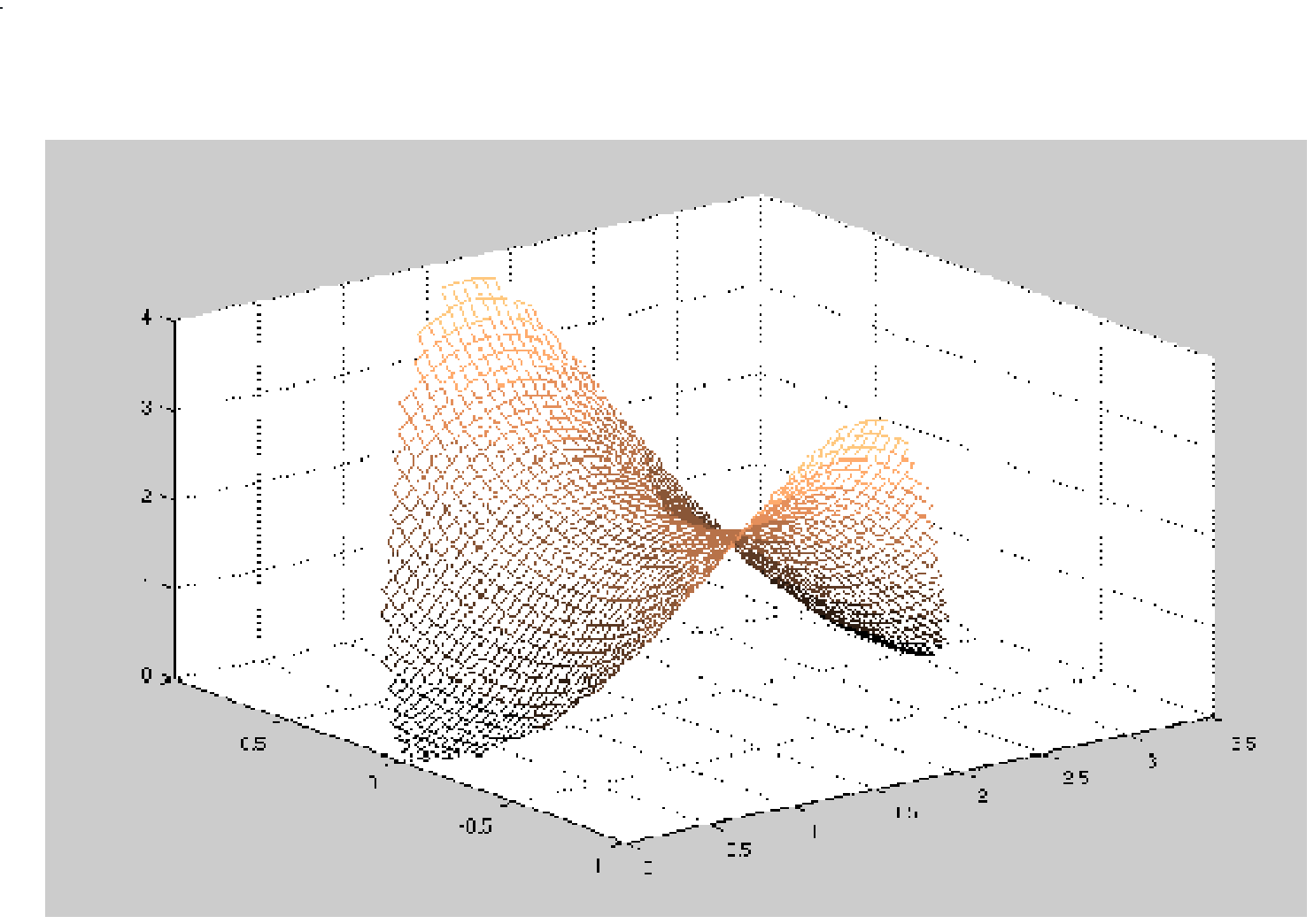}} 
\caption{ The extended probability density $\rho(x_r,x_i)$ for the infinite square well potential with $n=1$. }  \label{fig:infwell}}    \end{figure}

\begin{figure}[ht] 
\centering{\resizebox {0.8 \textwidth} {0.8 \textheight }  
{\includegraphics {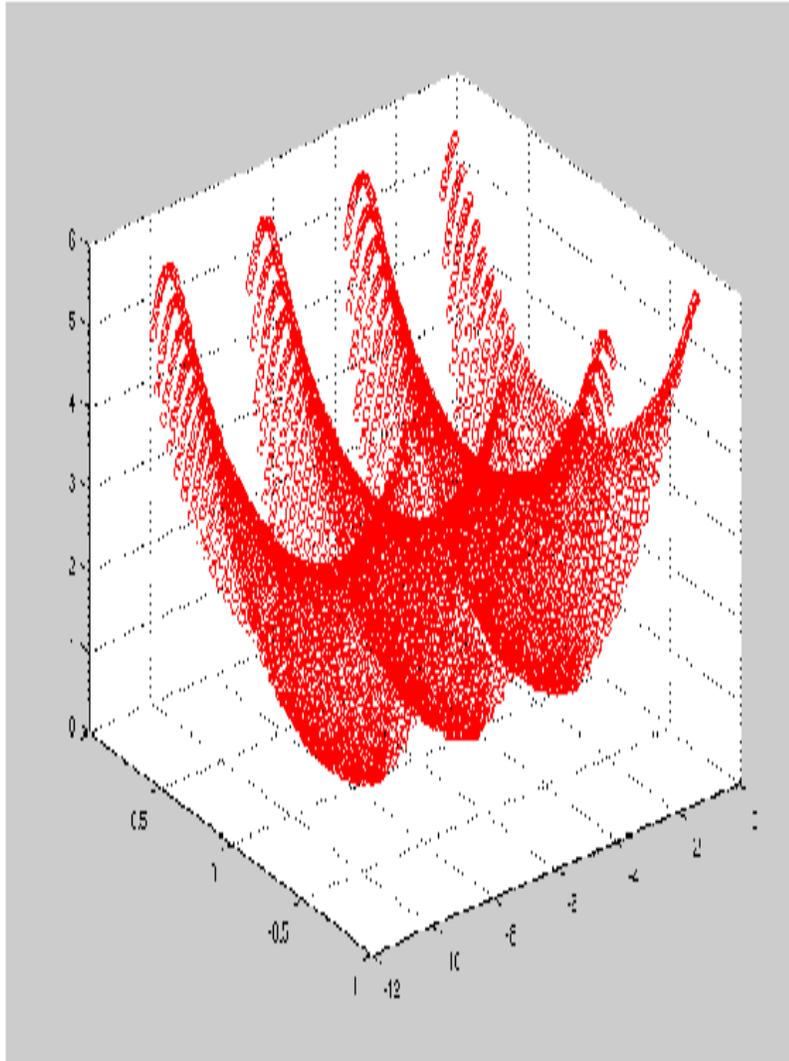}} 
\caption{  The extended probability density $\rho(x_r,x_i)$ for the potential step problem in the region where $V=0$.}  \label{fig:step}}    \end{figure}

\end{document}